\documentclass[a4paper,11pt]{article}
\usepackage{pos}
\usepackage{gensymb}
\usepackage{textgreek}

\usepackage[utf8]{inputenc}
\usepackage{graphicx}
\graphicspath{{images/}}
\begin{document}
\title{FASER Tracker Detector -- Commissioning, Installation, and Functionality}
\manuallySeparateAuthors
\author*[a]{Savannah Shively}
 \author{, on behalf of the FASER Collaboration}
\affiliation[a]{Department of Physics and Astronomy, University of California Irvine,\\ 4129 Frederick Reines Hall, Irvine, California, United States of America}
\emailAdd{sshively@uci.edu}
\abstract{FASER (ForwArd Search ExpeRiment) fills the axial blindspot of other, radially arranged LHC experiments. It is installed 480 meters from the ATLAS interaction point, along the collision axis. FASER will search for new, long-lived particles that may be hidden in the collimated reaction products exiting ATLAS. The tracking detector is an essential component for observing LLP signals. FASER's tracking stations use silicon microstrip detectors to measure the path of charged particles. This presentation is a summary of one of FASER's latest papers "The tracking detector of the FASER experiment", which describes the functionality, construction and testing of the tracker detector. FASER is currently installed in the LHC, where it is ready for data collection.}
\FullConference{International Conference on High Energy Physics\\
Bologna (Italy)\\
6-13 July 2022}
\maketitle

\section{Introduction}
FASER was proposed in 2017~\cite{Feng:2017uoz}  to search for long-lived particles in the far-forward direction of collisions in ATLAS. The experiment was approved by CERN in 2019 and began taking cosmics data in 2021. It is optimized for Beyond Standard Model searches and is additionally capable of neutrino measurements with the addition of its sister project, FASERnu. Hypothesized BSM particles, such as the dark photon, are predicted to decay in FASER's first 1.5 meter magnet. The tracking system measures the path of their products: charged pairs of leptons. FASER is now taking data in Run 3.

\section{Tracking Detector}

FASER's tracking system has several tiers of complexity. Its most fundamental components are the \textbf{silicon microstrip modules}, donated by ATLAS. These semi-conductor trackers (SCTs) detect the passage of a charged particle and are read out by an application-specific integrated circuit (ASIC). Each module has two sides of parallel-strip detectors, which are glued together with a 40 mrad stereo-angle. This provides a spatial resolution of 17 \textmu m in the precision coordinate (perpendicular to the strips) and 580 \textmu m in the non-precision coordinate. One NTC thermistor on each side measures the temperature of each side of each module ~\cite{FASER:Tracker}. Eight modules  compose a tracking plane.

The \textbf{tracking planes} are aluminum frames that support the SCT modules and encloses them in a humidity- and temperature-controlled environment. The modules interface with the outside of the plane through flexible printed circuit boards (AKA pigtails) which are connected to one of the patch panels affixed to either side of the plane. The plane is connected electrically to FASER's power and readout systems. The planes have carbon-fiber covers over the region that would include the center of the incoming beam. There are three planes in every tracking station.

A \textbf{tracking station} contains three tracker planes suspended by an  aluminum scaffolding. FASER's spectrometer contains three of the total four tracking stations. The remaining tracking station is the most recently installed and the most upstream. This "interface tracker" improves the neutrino track reconstruction capabilities of FASER and FASERnu. The planes are staggered vertically inside the tracking station, ensuring that all charged particle tracks will have at least two space-point coordinates per station. 

Finally, the 3 downstream stations are supported by a connecting \textbf{backbone}. It serves as a reference structure for aligning the trackers and facilitates easy handling and transportation. A clamp system enables some angular adjustment with respect to the magnets.

\begin{figure}[h]
\centering
\begin{tabular}{@{}c@{}}
    \includegraphics[width=.3\linewidth]{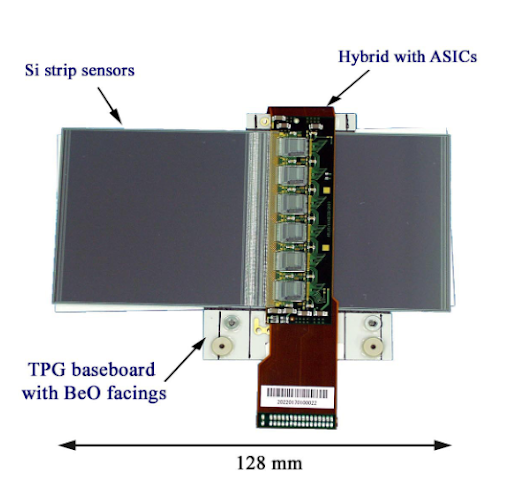} 
 \end{tabular}
 \begin{tabular}{@{}c@{}}
    \includegraphics[width=.5\linewidth]{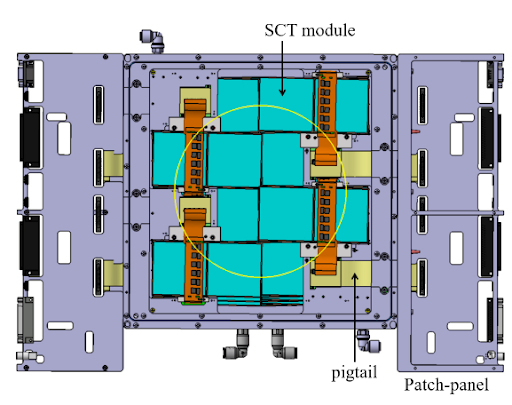} 
 \end{tabular}
 \begin{tabular}{@{}c@{}}
    \includegraphics[height=120pt]{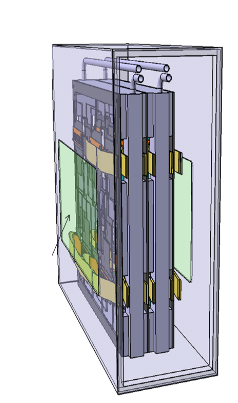} 
 \end{tabular}

\caption{(left) SCT Module. (center) Tracker plane, with eight SCT modules. (right) Tracker Station with three tracker planes.}
\label{fig:image2}

\end{figure}

\section{Monitoring and Protection}

\begin{figure}
    \centering
    \includegraphics[width=.7\linewidth]{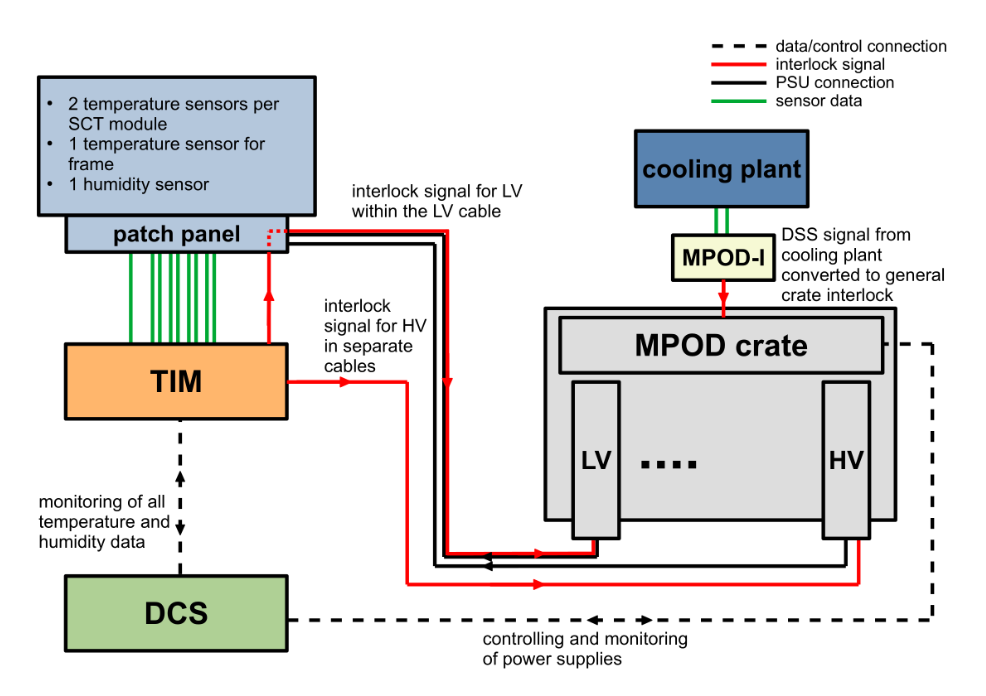}
    \caption{Schematic view of safety and interlock system.}
    \label{fig:interlock}
\end{figure}

The tracking system faces its greatest dangers from condensation and overheating. Temperature, humidity, and a derived dew point are observed through the Detector Control System (DCS), which receives data from Tracker Interlock and Monitoring Board (TIM) and can be checked from anywhere with access to the internet through grafana. 

The tracking planes use a thermal paste and a cooling system to keep module temperatures below 30\degree C. Cold (15\degree C) water is circulated through each tracking plane frame. Each station is flushed with dry air, which keeps the internal dew point below -40\degree C. If either of these systems fail, FASER's interlock system will warn monitoring personnel by email and Mattermost alert, and shut down the detector should the values reach a critical threshold. 

\section{Commissioning}
Commissioning the tracker system was conducted in stages corresponding to the tiers described earlier, beginning with the modules. The SCT modules were selected from the spares of those produced for ATLAS in 2004 which were still operational up to 300 V. The modules with the least defects were selected for installation in FASER. Quality was quantified with Leakage Current-Voltage (IV) scans at each stage of assembly.

The aluminium frames for the tracking planes were surveyed for precision after machining. Plane commissioning also included various monitoring and electrical tests to ensure good performance, stability and operation prior to installation in TI12. 

\section{Summary}
FASER is currently installed in TI12 and is taking data during Run 3. As of writing this proceedings, it has collected enough data to discover the dark photon within certain regions of coupling vs mass. FASER's tracker detector will be the primary tool for collecting data that would show evidence of BSM particles. In coordination with FASERnu, tracker data will help separate neutrino and anti-neutrinos and constrain their cross sections at TeV energies.~\cite{FASER:2019dxq}

\begin{figure}
    \centering
    \includegraphics[width=.7\linewidth]{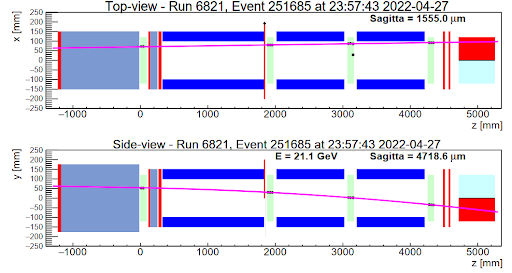}
    \caption{TI12 event display showing a muon track traversing all four FASER tracking stations.}
    \label{fig:ti12event}
\end{figure}

FASER has already seen its first tracks in TI12 from cosmic ray studies and has detected muons from collisions in ATLAS. Analysis efforts are ongoing to prepare all necessary tools to discover BSM physics and carry out neutrino studies. FASER expects to collect an integrated luminosity of as much as 300 fb\textsuperscript{-1} during Run 3. 

\section{Acknowledgements}
This work was supported in part by U.S. National Science Foundation Grants PHY-2111427 and PHY-2210283, Simons Foundation Grant 623683, and the Corwin and Nancy Evans Award.

\bibliography{main}

\end{document}